# THE ORIGIN OF GALACTIC COSMIC RAYS

Peter L. Biermann,[a] T.K. Gaisser[b] & Todor Stanev[b]

[a]Max-Planck Institute for Radioastronomy,

Bonn, Germany

[b]Bartol Research Institute, University of Delaware,

Newark DE 19716

## Abstract

Motivated by recent measurements of the major components of the cosmic radiation around 10 TeV/nucleon and above, we discuss the phenomenology of a model in which there are two distinct kinds of cosmic ray accelerators in the galaxy. Comparison of the spectra of hydrogen and helium up to 100 TeV per nucleon suggests that these two elements do not have the same spectrum of magnetic rigidity over this entire region and that these two dominant elements therefore receive contributions from different sources.

# 1 Introduction

Recent measurements of the flux of helium at high energy[1, 2] show that its spectrum is different from that oftable protons [3, 4, 5]. In particular, there is no hint in the helium spectrum of the steepening that appears to be present in the proton spectrum at a rigidity of about 40 TV [1]. To the extent that acceleration and propagation of cosmic rays depend on collisionless processes, such as acceleration at supernova blast waves and diffusion in turbulent plasmas, the particle spectra at high energy should depend only on gyroradius, or equivalently on magnetic rigidity. Thus a difference between the rigidity spectra of protons and helium would require different acceleration sources and/or propagation histories.

Webber & Lezniak [6] showed that, below 100 GV the rigidity spectra of hydrogen and helium are consistent with each other within experimental errors. The observed number ratio of 7:1 for cosmic ray hydrogen:helium in this range of rigidities corresponds to a source ratio that is somewhat lower than the general abundance ratios of these elements [7]. The natural conclusion from these observations was that these two species have the same source and propagation histories. Present evidence suggests that this is not the case over the larger energy range up to 100 TeV/nucleon.

In Fig. 1 we show the rigidity spectra of hydrogen and helium above 10 GV from several different experiments. No single experiment spans the whole energy region, so conclusions about the overall shape of the spectra must be qualified by the problem of systematic effects, which can shift the normalization of one experiment relative to another. Nevertheless, inspection of Fig. 1 suggests that the spectrum of helium is somewhat flatter than that of protons even before the steepening of the proton spectrum.

It is also interesting to compare the spectral indices reported by the various experimental groups in their limited energy ranges, which we do in Table 1. All these spectral indices are consistent with each other within their stated errors, but they are also consistent with a small difference between the spectrum of hydrogen and helium over the whole energy range. Because of the systematic problem referred to above, it is impossible to tell from the data alone whether helium has a different spectrum from hydrogen already at low energy or whether the difference occurs only in the JACEE energy range above $10^4$ GV. The strongest evidence for a difference between hydrogen and helium comes from within the JACEE data. The difference in their fits to the proton spectra above and below 40 TeV corresponds to a statistical significance of approximately $3\sigma$ for the steepening of the proton spectrum. In contrast, the JACEE helium spectrum shows no sign of a steepening over this energy range. In addition, it fits smoothly onto the RICH helium spectrum [2] at lower energy.

Table 1. Spectral indices for hydrogen and helium.

| Experiment | Energy Range (p) | Hydrogen | Helium |
|---|---|---|---|
| Webber [8] | 1–50 GeV | $2.70 \pm 0.05$ | $2.70 \pm 0.05$ |
| LEAP [9] | 10–100 GeV | $2.74 \pm 0.02$ | $2.68 \pm 0.03$ |
| Ryan et al. [10] | 50–2000 GeV | $2.75 \pm 0.03$ | $2.77 \pm 0.05$ |
| JACEE [1, 3] | 50–200 TeV | $2.77 \pm 0.06$ | $2.67 \pm 0.08$ |
| JACEE [3] | <40 TeV | $2.64 \pm 0.12$ | — |
| JACEE [3] | >40 TeV | $3.22 \pm 0.28$ | — |
| RICH [2] | 100–1000 GV | — | $2.64 \pm 0.09$ |
| Sokol [5] | >5 TeV | $2.85 \pm 0.14$ | $2.64 \pm 0.12$ |
| MSU [4] | 10–200 TeV | $3.14 \pm 0.08$ | — |
| Japan [11] | 8–50 TeV | $2.82 \pm 0.13$ | $2.75 \pm 0.15$ |

## 2 Overview of spectra

The new high energy data continues a trend that has been observed by several different experiments – heavy nuclei (possibly including helium) have slopes flatter than the canonical 2.75 observed for protons up to 1000 GeV. To demonstrate this trend, we consider in this section a simple two-component fit to the spectra of five different groups of nuclei: hydrogen, helium, carbon–oxygen, neon–silicon and iron. These fits are extrapolations to low energy of the model of Biermann et al. [12, 13, 14, 15], which we review briefly in the next section.

Fig. 2 shows the proton and helium energy spectra above $\sim$ 10 GeV/nucleon. The measured differential flux is multiplied by $E^{2.75}$ to flatten the spectrum. This factor however exaggerates the uncertainty in the energy determination and translates small systematic errors in the energy assignment into normalization uncertainties on the plot.

The dotted lines in Fig. 2 show $E^{-2.75}$ spectra with an exponential cutoff at $1.2\times10^5$ GV, and the dashed lines represent $E^{-2.67}$ spectra with a steepening at $7\times10^5$ GV [15]. The solid line is the sum of the two components. A slightly flatter second component would fit better the low energy normalization of Seo *et al.*[9] and Webber et al.[8], but would decrease the agreement with the RICH [2] data.

Fig. 3 shows the spectra of heavier nuclei, divided into three groups: C–O, Ne–Si and Fe. The grouping is necessary for the extension of the spectra to high energy because of the decreasing charge resolution and the low fluxes, and correspondingly the low experimental statistics at high energy. The C–O data of HEAO [16] and CRN [17] are the sums of carbon and oxygen, while the JACEE statistics [1] contain some nitrogen nuclei. For heavier nuclei the HEAO and CRN points are the sums of the Ne, Mg and Si fluxes, while JACEE data are for all nuclei between Ne and S. The JACEE data for the Fe group may contain nuclei of the sub–iron group. We only present data for energy above 10 GeV/nucleon to avoid the need to account for the solar modulation. The solid curves in Fig. 3 show only the flatter($\propto E^{-2.67}$) component, since the steeper one is not needed for the heavier nuclei. The slight curvature of the lines reflects details in the model of Refs. [12, 13, 14], which only have small effects in the energy range under discussion here. We note that the recent Fe data of Ichimura *et al.* [18] require a slightly higher normalization and/or flatter spectral index than the other data sets.

For heavy nuclei, some flattening of the spectrum is expected from effects of propagation, which are not included in the solid curves of Fig. 3. We estimate the size of this contribution within the context of the "leaky box" model, which is adequate for this purpose. If galactic cosmic ray sources accelerate primary cosmic ray nuclei (e.g. Fe) at the rate $Q_{\rm Fe}(R)$ particles cm$^{-3}$s$^{-1}$GeV$^{-1}$, and escape of particles from the Galaxy is char-acterized by a rigidity-dependent time, $\tau_{esc}(R)$, then the observed cosmic ray intensity at a typical location inside the propagation volume (e.g. in the local interstellar medium) will be

$$\Phi_{Fe} = \frac{\frac{c}{4\pi} Q_{Fe}(R) \tau_{esc}(R)}{1 + \lambda_{propagation}(R)/\lambda_{Fe}}. \quad (1)$$

Here $\lambda_{Fe} \approx 2.6$ g/cm$^2$ is the interaction length for iron in the interstellar medium and $\lambda_{propagation}$ is the amount of matter that a particle encounters on propagation from the source to Earth. Following Ref. [16], we assume that $\lambda_{propagation} \equiv \lambda_{esc} = \beta c \rho \tau_{esc} \propto R^{-\delta}$ and normalize $\lambda_{esc} = 5.4$ g/cm$^2$ at $R = 21.5$ GV, which corresponds to an energy of 10 GeV/nucleon for iron. The differential source spectrum is $Q \propto R^{-\alpha}$. For protons and light nuclei, for which $\lambda_{esc}/\lambda_{interaction} \ll 1$ in Eq. 1, at energies of interest here, the observed differential spectral index is $-(\alpha + \delta)$. When $\lambda_{esc}/\lambda_{interaction}$ is not small the observed spectrum is flatter, approaching the source spectrum for heavy nuclei at low energy.

In the model of Ref. [12], $\delta = 1/3$, which follows from the energy-dependence of diffusion in a medium characterized by a Kolmogorov spectrum of turbulence [19]. The source spectrum in that model is $\alpha = 7/3$. The dashed line for iron in Fig. 3 is a plot of Eq. 1 with these parameters. A more conventional set of parameters (e.g. Ref. [16]) would require $\alpha \approx 2.1$ and $\delta \approx 0.6$. This is shown by the dotted line in Fig. 3. The curves are all normalized at 55 GeV/nucleon. At high energy $\lambda_{esc} \ll \lambda_{Fe}$, and all three curves approach $\alpha + \delta = 2.67$. The errors of the experimental points are quite large and do not allow us to fix the normalization of the Fe source flux better than 30 – 40%. Either of these propagation models suggests that iron is nearly a factor of two more abundant "at the source" than measured at 50 GeV/nucleon.

For lighter nuclei, the effect of propagation will be smaller than illustrated here for iron. In view of the fact that there is freedom to renormalize the source spectrum as $\delta$ changes, we conclude that there is no need to introduce the steeper component to fit the data for the nuclei in Fig. 3. The most abundant nuclei, represented by the five groups in Figs. 2 and 3, can be associated with two classes of sources. The proton flux is dominated by source I, with a spectral index of $\approx 2.75$ at Earth. The heavy nuclei (including most helium) are accelerated at source II with an index of $\sim 2.67$ after accounting for propagation.

We emphasize that any conclusion based on the simple fits described above is at best suggestive. This is because there is no theoretical basis for the use of a single power to extrapolate the high energy fits to low energy. On the contrary, there are reasons that a single power should not be correct, which we discuss in the next section. In addition, as shown in Ref. [17], the data for the heavy nuclei themselves indicate some differences among spectral indices beyond what arises from propagation in a leaky box model. The exercise illustrated in Figs. 2 and 3 does demonstrate, however, that it is possible that helium and hydrogen may have different sources even at low energies $\sim 10$ GeV/nucleon, and raises the possibility that the sources of cosmic ray helium may be more closely related to those of heavier nuclei rather than to the sources of protons.

## 3 A possible model

Starting with the concept that supernova explosions into stellar winds reproduce the abundances of such winds, and thus represent an enriched component of source gas for cosmic ray acceleration [20], Silberberg *et al.* [21] produced an estimate for the expected cosmic ray abundances from such sources. In making their estimate they accounted for the relative abundances of various types of supernova progenitors and the properties of the winds of the massive progenitors. They also included the effect of the first ionization potential (FIP) on the injection of various elements into the acceleration process. These ideas have been further elaborated in a series of papers by Biermann and collaborators [12, 13, 14].

The basic premise of this theory is that galactic cosmic rays originate from two different sites, 1) Sedov type supernova explosions into the interstellar medium, and 2) supernova explosions of massive stars into their own stellar wind. The theory makes specific predictions for the spectral index of the wind component below and above the knee, as well as for the spectral index of the Sedov component. The cutoff of the Sedov–component and the location of the knee feature[1] are also predicted and checked against a variety of observations. The comparison with the shower size spectra in the region of the "knee" ($\sim 3 \times 10^{15}$ eV) made in Ref. [15] and the cosmic ray composition above the knee [22] showed that the parameters of the model, when fitted to these data, were in reasonable agreement with the values predicted [23]. Here we wish to identify the two groups introduced above with these two source sites.

Wheeler [24] discusses supernova rates and stellar evolution. He notes that supernovae of type Ia are likely to be explosions of white dwarfs, while the other supernova types probably all are from originally massive stars, above a zero-age main sequence mass of about 8 $M_\odot$. The supernova rates determined for galaxies similar to our own are subject to a number of important selection effects, but the numbers indicate at present that supernovae of type Ia are only about 10% of all supernovae. Wheeler also notes that mass loss becomes important for zero age main sequence stars above 15 $M_\odot$. This mass loss arises in the form of strong winds, so we can tentatively identify the mass range above 15 $M_\odot$ with those supernova events which give rise to the wind component. The supernovae between 8 $M_\odot$ and 15 $M_\odot$ plus those of type Ia give the Sedov component.

---

[1] The change in spectral slope of the cosmic ray spectrum at the "knee" is attributed to a reduction in acceleration efficiency at a specific rigidity [12].

Using the observed mass distribution of stars and assuming that all supernovae produce approximately the same energy in cosmic rays, one can estimate the energy ratio between the two kinds of cosmic ray sources. ¿From Wheeler's Fig. 10, the ratio of supernovae above and below 15 $M_\odot$ is about 1 to 3. The addition of white dwarf supernovae, a 10% effect, does not change this result. On the other hand, integrating the energy contained in the cosmic rays of the two source sites, directly from the graphs in Ref. [15] or from the graphs in this paper, gives also a ratio very close to 1/3, as already discussed by Biermann and Cassinelli[13]. The supernova rates expected from the statistics of supernova events in galaxies similar to our own thus provide just the numbers needed to understand the energy of cosmic rays from the two different types of sources, Sedov supernovae and wind-supernovae.

## 4  Discussion and Conclusions

Other authors [25, 26] have also realized the necessity for injection of material enriched in heavy elements to explain the cosmic ray abundances. The model discussed above, without any additional assumptions, accounts well for the underabundance of hydrogen and helium relative to silicon at low energies. In this scenario, hydrogen is underabundant because it comes from the Sedov–type explosions into the interstellar medium and silicon comes from wind explosions. Helium is underabundant because the winds of massive stars, i.e. blue and red supergiants as well as Wolf Rayet stars, are enriched in heavy elements, as discussed by Silberberg *et al.* [21].

An alternative is that the precise relative weights of the two components for helium are different from the Biermann *et al.* model and the second component only dominates for helium at high energy (e.g. >TeV/nucleon). The weakness of this alternative is that it would require fine tuning to produce the smooth helium spectrum.

Development of a complete model based on the two-source scenario described above would require that several aspects be treated in a much more realistic way than we have attempted here. For example, one would need to account for the locations of the various kinds of supernovae in the Galaxy and the sizes of the astrospheres created by the progenitor winds in the case of different sizes of massive stars. One would also need to follow the time history of the acceleration as in Ref. [19]. It is possible that only the flat, high energy part of the spectrum would be produced during the expansion of the SN blast wave through the progenitor wind.[2] The steeper, low energy part of the spectrum might be Sedov-like, produced after the blast wave breaks out of the progenitor's astrosphere. The time dependence of the development of the cosmic ray spectrum produced by an expanding supernova blast wave, convoluted with characteristic distributions of various elements around the progenitors, could lead to an interesting and complicated spectrum for each element. Combining this with propagation effects, which would be different for different classes of supernovae to the extent that they have different distributions in the Galaxy, would complicate the situation still further.

As mentioned earlier, the models of Biermann and collaborators use an escape probability with an energy dependence of $E^{-1/3}$. This theoretically motivated value is not in direct contradiction with the $E^{-0.6}$ energy dependence, derived from the measurements of the secondary to primary nuclei ratio in cosmic rays[16]. The secondary/primary ratio measures the amount of matter traversed by cosmic rays, and could be strongly influenced by the matter distribution in the galaxy and its temporal behaviour. The derivation of the secondary/primary ratio also depends crucially on the energy behaviour of the spallation cross–sections. One recent study[27] suggests that a better representation of the energy dependence of these cross–sections may decrease significantly the pathlength dependence on rigidity. Another point to note from Fig. 3-c is that the flattening between 10 and 100 GeV/nucleon may not be as great in the data as in the leaky box model, especially when $\delta \sim 0.6$. This point was noted in Ref. [17].

---

[2]This is perhaps what motivated Silberberg *et al.*[21] to attribute only the very high energy particles to the wind supernovae in the first place.

Among the consequences of a model in which hydrogen and helium come from different kinds of sources is that the propagation parameters derived from one of the populations can not be applied to the other. The possible difference in the propagation history of light and heavy cosmic ray population may change, for example, the estimates of the GeV antiproton fluxes. In addition, if the low energy helium is also from the wind component, the correlation [28] between the ratio $He^3/He^4$ and $\bar{p}/p$ would also break down.

**Acknowledgements.** The authors thank J.P. Wefel, E.-S. Seo and W.R. Webber for sharing data in tabular form with us. We are grateful to J.P. Wefel and E.-S. Seo for helpful discussions and to Gary Zank for comments on an earlier version of this paper. We thank a referee for pointing out a numerical error in our original Fig. 3. The research of T.S. and T.K.G. is funded in part by DOE and NASA and that of P.L.B. by DFG Bi 191/9 and BMFT DARA 50OR9202. We also share NATO CRGP grant 910072.

**Figure Captions.**

Fig. 1. Hydrogen and helium rigidity spectra above 10 GV: The open circles are from Webber *et al.* [8], triangles are from Ref. [10], and inverted triangles from LEAP[9]. Dots represent the measurements of JACEE[3, 1], squares – RICH[2], and the crosses – Kawamura *et al.* [11].

Fig. 2. Hydrogen and helium energy spectra above 10 GeV/nucleon. The data points are indicated as in Fig. 1. Dotted lines show the contribution of Source I (Sedov–type explosions), dashed lines – that of Source II (wind–explosions). The solid lines are the sum of both sources.

Fig. 3. Spectra of heavy nuclei above 10 GeV/nucleon. The triangles are from HEAO[16], the squares – from CRN[17], the dots – from JACEE[1] and the crosses – from Ichimura *et al.* [18]. The solid lines show the predicted spectra from Ref. [15]. The broken lines illustrate the effects of propagation on the observed spectrum of iron (see text).

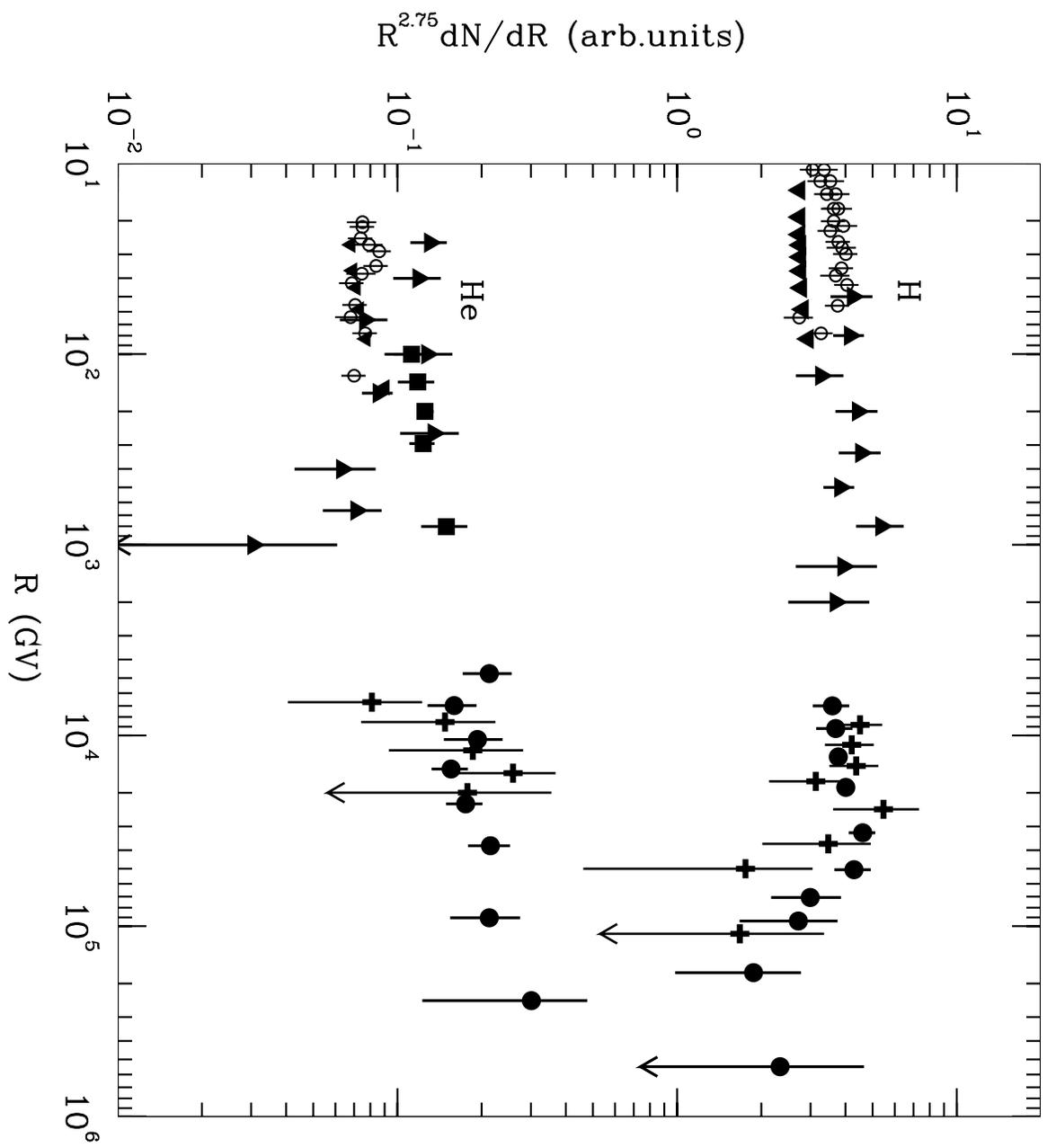

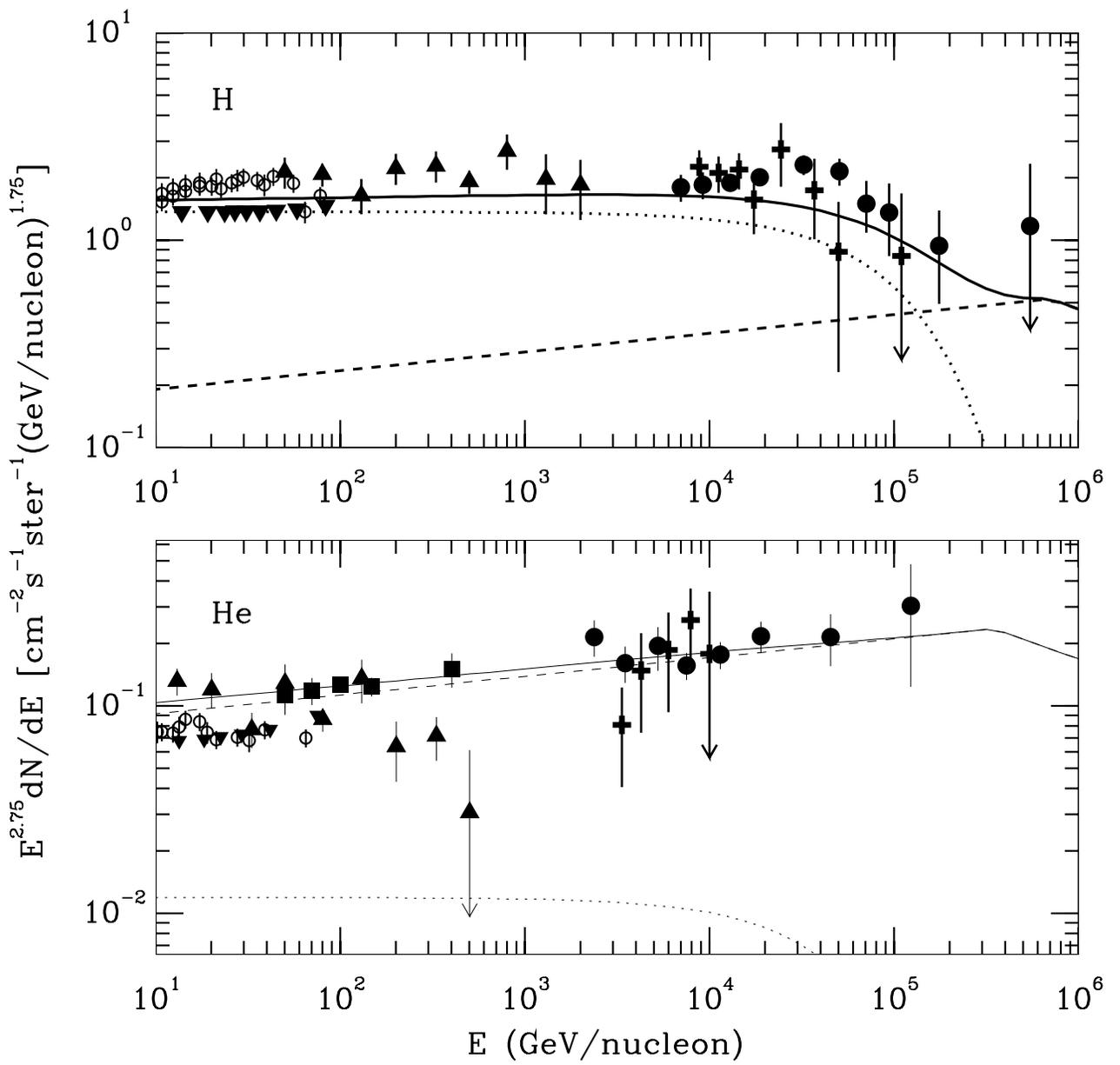

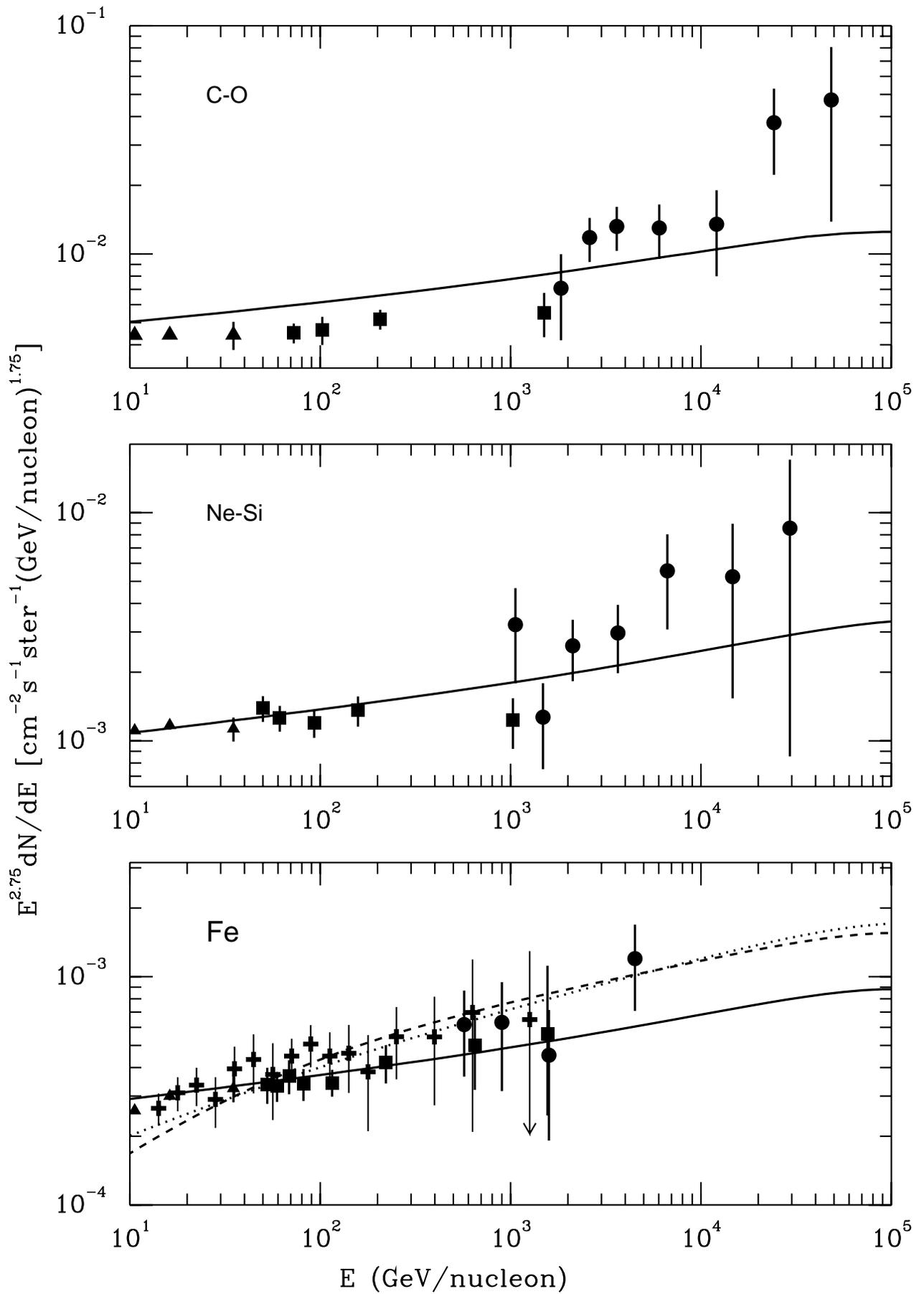